\documentclass[nameyear,preprint,12pt,authoryear]{elsarticle}
\usepackage{rotating}
\usepackage{lineno,hyperref}
\modulolinenumbers[5]
\journal{Journal of \LaTeX\ Templates}

\usepackage[varg]{txfonts}
\usepackage{times}

\usepackage{multirow,graphicx,bm,amssymb}
\usepackage[normalem]{ulem}
\usepackage{soul}
\usepackage{epsfig}
\usepackage{natbib}
\journal{New Astronomy}
\usepackage{psfrag}

\def\apj{ApJ}
\def\apjl{ApJ}

\def\apss{Ap\&SS}
\def\aap{A\&A}

\def\mnras{MNRAS}

\def\na{New A}

\def\prd{Phys.~Rev.~D}

\def\pasp{PASP}

\def\nat{Nature}

\def\physrep{Phys.~Rep.}

\newcommand{\msun}{\mbox{$M_\odot$}}
\newcommand{\rsun}{\mbox{$R_\odot$}}
\newcommand{\Mbh}{\mbox{$M_{\rm BH}$}}
\newcommand{\Mpns}{\mbox{$M_{\rm PNS}$}}
\newcommand{\cyg}{Cyg X-1~}
\newcommand{\lmc}{LMC X-1~}
\newcommand{\mx}{M33 X-7~}

\def\be{\begin{eqnarray}}
\def\ee{\end{eqnarray}}
\def\bi{\begin{itemize}}
\def\ei{\end{itemize}}
\def\lsim{\mathrel{\rlap{\lower3pt\hbox{\hskip1pt$\sim$}}
     \raise1pt\hbox{$<$}}}
\def\gsim{\mathrel{\rlap{\lower3pt\hbox{\hskip1pt$\sim$}}
     \raise1pt\hbox{$>$}}}

\hypersetup{colorlinks=false,pdfborder={0 0 0},}

\begin{document}

\begin{frontmatter}

\title{Limits On The Spin Up Of Stellar-Mass Black Holes Through A Spiral Stationary Accretion Shock Instability}


\author{Enrique Moreno M\'endez}
\address{Instituto de Astronom\'ia, Universidad Nacional Aut\'onoma de M\'exico, Circuito Exterior, Ciudad Universitaria, Apartado Postal 70-543, 04510, D.F., M\'exico.}
\ead{enriquemm@ciencias.unam.mx} 

\author{Matteo Cantiello}  
\address{Kavli Institute for Theoretical Physics, University of California, Santa Barbara, CA 93106, USA.} 
\ead{matteo@kitp.ucsb.edu}

\begin{abstract}
The spin of a number of black holes (BHs) in binary systems has been measured. In the case of BHs found in low-mass X-ray binaries (LMXBs) the observed values are in agreement with some theoretical predictions based on binary stellar evolution.  However, using the same evolutionary models,  the calculated  spins of BHs in high-mass X-ray binaries (HMXBs) fall short compared to the observations.
A possible solution to this conundrum is the accretion of high-specific-angular-momentum material after the formation of the BH, although this requires accretion above the Eddington limit.
Another suggestion is that the observed high values of the BHs spin  could be the result of an asymmetry during Core Collapse (CC).
The only available energy to spin up the compact object during CC is its binding energy.
A way to convert it to rotational kinetic energy is by using a Standing Accretion Shock Instability (SASI), which can develop during CC and push angular momentum into the central compact object through a spiral mode ($m = 1$).
Here we study the CC-SASI scenario and discuss, in the case of  LMXBs and HMXBs, the limits for the spin of a stellar-mass BHs.
Our results predict a strong dichotomy in the maximum spin of low-mass compact objects and massive BHs found in HMXBs.
The maximum spin value ($|a_\star|$) for a compact object near the mass boundary between BHs and NSs is found to be somewhere between 0.27 and 0.38, depending on whether secular or dynamical instabilities limit the efficiency of the spin up process.
For more massive BHs, such as those found in HMXBs, the natal spin is substantially smaller and for $\Mbh\!>\!8~\msun$  spin is limited to values $|a_\star|\!\lesssim\!0.05$.
Therefore we conclude that the observed high spins of BHs in HMXBs cannot be the result of a CC-SASI spin up.

\end{abstract}

\end{frontmatter}
  

\section{Introduction}
\label{sec-Intro}

It is generally accepted that stellar-mass Black Holes (BHs), like neutron stars (NSs), are formed by the core collapse (CC) of a massive star
($M\gtrsim8\msun$) \citep[][]{1983bhwd.book.....S}.
Whether a BH or a NS is formed depends mostly on whether the maximum stable mass of a NS is exceeded during CC or shortly after by fall back material.
The mass cutoff between BHs and NSs is still not well known, but it is generally accepted to lie somewhere between 2.5 and $3~\msun$ \citep[][]{2010arXiv1012.3208L}.

If a NS is produced and its spin period is short enough ($P\sim10$ s to $100~$ms) a pulsar (PSR) is born.
Theory still has difficulties explaining the observed spin rates.
From the stellar-evolution point of view, models of massive stars that include angular-momentum transport due to rotationally-induced instabilities and circulations, over-predict the final spin of NSs \citep[e.g.,][]{2000ApJ...528..368H}.
The situation changes in models that also include transport of angular momentum by magnetic torques, which do a much better job in predicting the final rotation rate of NSs \citep{2005ApJ...626..350H,Suijs_2008}. 
Note however that the physics of internal angular momentum transport in stars is not yet understood \citep{Cantiello_2014,Fuller_2014}, and the spin rate of pre-collapse stellar cores might be determined by processes not currently included in stellar evolution calculations \citep{2015arXiv150207779F}.  
Regardless, the final spin of a NS can be further affected by asymmetries of the CC and/or supernova (SN) explosion \citep[see, e.g.,][for recent simulations]{2013A&A...552A.126W}.
\citet{1998Natur.393..139S} proposed that during the SN explosion, an asymmetric  kick (off the radial direction of the star), or a series of kicks, could be responsible for spinning up NSs as well as for giving them large radial velocities.
More recently, \citet{2007Natur.445...58B} discussed a mechanism that may allow NSs to be spun up during stellar-core collapse.
Said mechanism relies on an instability of the accretion shock (Standing Accretion Shock Instability, SASI) which is formed after the bounce of the core during the stellar collapse. A spiral mode of this instability can lead to a spin-up of the PNS \citep{2000A&A...363..174F,2002A&A...392..353F,2007ApJ...654.1006F,2008A&A...477..931S,2010ApJ...725.1563F,
2011ApJ...732...57R,2013ApJ...770...66H,2015MNRAS.452.2071F}.
In this situation even an originally non-rotating progenitor could, in principle, form a rapidly spinning NS.

In this paper we study the SASI spin-up scenario and determine upper limits for the natal  spin of compact objects by assuming all the available (binding) energy is transformed into kinetic rotational energy.
Our results show that this mechanism (essentially, conservation of energy and angular momentum during CC) can not explain the high spins of BHs in HMXBs.

\subsection{Black Holes' Spin}
\label{subsec-KBHs}

\begin{sidewaystable*}
\label{table:bhs}
\begin{center}
\caption[]{Physical parameters for six BH binaries (three LMXBs and three HMXBs) listed in order of increasing $a_\star$.
For each system we show mass of the black hole and the companion, orbital period as well as predicted and measured Kerr parameter $a_\star$.
REFERENCES: (1) \citet{2011MNRAS.tmp.1036S}, (2) \citet{2011ApJ...730...75O}; (3) \citet{2006ApJ...636L.113S}, (4) \citet{2002MNRAS.331..351B};
(3) \citet{2006ApJ...636L.113S}, (5) \citet{2004ApJ...610..378P}; (6) \citet{2011arXiv1106.3688R}, \citet{2011arXiv1106.3689O} and \citet{2011arXiv1106.3690G} who have new estimates of distance (from trigonometric parallax), mass and spin, however, compare to (7) \citet{2005AA...438..999A} who find $0.48\pm0.01$ or (8) \citet{2009ApJ...697..900M} who find $a_\star\!=\!0.05$; previous estimates for masses can be found at (9) \citet{1995A&A...297..556H} and (10) \citet{2003IAUS..212..365O}; (11) \citet{2008ApJ...679L..37L}, (12) \citet{2010ApJ...719L.109L}, (13) \citet{2007Natur.449..872O}, (14) \citet{2004AA...413..879P}; (15) \citet{2009ApJ...701.1076G}, (16) \citet{2009ApJ...697..573O}.}\label{Tab:Binaries}
\begin{tabular}{lcccccc}
\hline\hline
    BH Binary       &        $\Mbh$          &        $M_{\rm sec}$        &    $P_{\rm now}$    &  $a_\star$  &        $a_\star$       &     References   \\
                    &       $[\msun]$        &        $[\msun]$        &     [days]      &   (Model)   &        (Measured)      &                  \\
\hline\hline
\multicolumn{7}{c}{LMXBs} \\
\hline
    XTE J1550-564 &     $9.10\pm0.61$      &     $0.30\pm0.07$      &  $1.5420333(24)$ &    $0.5$    &  $0.49^{+0.13}_{-0.20}$ &    1,2,\\
    GRO J1655-40  &      $5.4\pm0.3$       &   $1.45\pm0.35$        &      $2.6127(8)$ &    $0.8$    &       $0.65-0.75$       &    3,4          \\
    4U 1543-47    &      $9.4\pm2.0$       &     $2.7\pm1.0$        &         $1.1164$ &    $0.8$    &       $0.75-0.85$       &    3,5          \\
\hline\hline
\multicolumn{7}{c}{HMXBs} \\
\hline
      \mx           &   $15.65\pm1.45$       &      $70.0\pm6.9$      &       $3.453014$ &    $<0.15$  &     $0.84\pm0.05$       &    11,12,13,14  \\
      \lmc          &   $10.91\pm1.41$       &    $31.79\pm3.48$      &     $3.90917(8)$ &    $<0.15$  &  $0.92^{+0.05}_{-0.07}$ &    15,16        \\
      \cyg          &   $14.81\pm0.98$       &      $19.2\pm1.9$      &       $5.599829$ &    $<0.15$  &    $>0.97(3\sigma)$     &    6,7,8,9,10   \\
\hline\hline
\end{tabular}
\end{center}
\end{sidewaystable*}

\citet{2002ApJ...575..996L}, \citet{2007ApJ...671L..41B} and \citet{2011ApJ...727...29M} have estimated the spin\footnote{$a_\star=Jc/GM^2$, where $J$ is the total angular momentum, $c$ is the speed of light, $G$ is the gravitational constant and $M$ is the mass of the BH.} of the Galactic, stellar-mass BHs where the masses and orbital periods of the binaries are relatively well constrained (some of them are included in Table~\ref{Tab:Binaries}).
In their model a massive star and its companion evolve into a common-envelope phase after Case-C mass transfer \citep[i.e., mass transfer taking place after core-He burning in the primary;][]{1970A&A.....7..150L} 
and the orbital separation decreases to a few $\rsun$ after the removal of the envelope of the primary star.
The mass of the companion constrains the separation of the binary at the time the BH forms, as a massive companion cannot fit in arbitrarily small orbits.
Following the formalism of \citet{1975A&A....41..329Z,1977A&A....57..383Z,1989A&A...220..112Z} it is assumed that the tidal-synchronization timescale of the massive star is short enough to allow for the star to rotate synchronously with the orbital period \citep[later confirmed numerically by][]{2007Ap&SS.311..177V}. 
More recently, \citet{2014ApJ...781....3M} estimated the Alfv\'en timescales for angular momentum transport in and out of the inner layers of the star, also finding ranges in the magnetic field that allow to keep a substantial amount of angular momentum in the stellar core.
Thus, the maximum spin the BH can have is constrained by tidal synchronization with the companion star at a very late stage in its evolution \citep[see, e.g., Fig.~3 in ][]{2008ApJ...685.1063B}.
This resulted in the predicted values for $a_\star$ shown in Table~\ref{Tab:Binaries}.
Notice that all of the predicted and measured $a_\star$s are positive, 
most likely as a consequence of tidal synchronization.
Besides those in table~\ref{Tab:Binaries}, there are three more with measured Kerr parameters.
GRS 1915$+$105, which has an $a_\star \gtrsim 0.98$ \citep[][; compare to the model prediction for natal spin: $a_\star \sim 0.2$]{2006ApJ...652..518M}; it is likely that the BH in this system accreted $\gtrsim 50 \%$ of its present mass (which is currently $\Mbh \sim 15 \msun$) via RLOF \citep[][]{2011ApJ...727...29M}.
A large mass transfer explains the present spin and its long orbital period (33.5 days).
A second system is A 0620-003, with $a_\star = 0.12 \pm 0.19$ \citep[][compare to the predicted spin $a_\star = 0.6$]{2010ApJ...718L.122G}; 
this system is harder to model given that the companion is a low-mass star ($\sim 0.7 \msun$). 
As the system may be very old, the long timescale available for angular momentum loss through different channels makes reliable estimates difficult.
The third system is LMC X-3, which has a measured spin of $a_\star \lesssim 0.3$ \citep{2006ApJ...647..525D},
and \citet{2008ApJ...685.1063B} estimate it may have been formed with $a_\star \simeq 0.5$ before powering up a long GRB/hypernova event.
Furthermore, this model along with that of \citet{1992Natur.357..472U}, has been recently employed to explain a three-peaked GRB (110709B) in \citet{2014arXiv1411.7377M}.

Measurements of the spins of several sources have confirmed the theoretical predictions of \citet{2002ApJ...575..996L}, \citet{2007ApJ...671L..41B} and \citet{2011ApJ...727...29M}  on three Galactic sources. These are GRO J1655-40, 4U 1543-47 \citep{2006ApJ...636L.113S} and more recently, XTE J1550-564 \citep{2011MNRAS.tmp.1036S}.
However, the predicted natal spins fall short with respect to the observed ones for BHs with massive companions (see last three rows in Table~\ref{Tab:Binaries}).
\citet{2008ApJ...689L...9M} have addressed this issue by arguing that later accretion onto the BH can spin it up.
\citet{2008ApJ...679L..37L}, \citet{2009ApJ...701.1076G} and later \citet{2010Natur.468...77V} as well as \citet{2010arXiv1011.4528A} argue against such a scenario based on the fact that this would require mass transfer rates above Eddington's limit.
They also note that due to the large mass ratio in such binaries, mass transfer would be unstable and quickly lead to a merger, thus preventing the observation of such systems.  
Hence, they argue for the spin of these BHs being natal.
\citet{2010arXiv1011.4528A} further conclude that in the case of \cyg the spin could not be aquired previously to the formation of the BH \citep[for similar reasons to those of][]{2007ApJ...671L..41B,2011ApJ...727...29M}, and hence, it must have been acquired during the collapse into the BH, 
and suggest this may have occurred through the mechanism put forward by \citet{2007Natur.445...58B} for spinning up NSs.
Meanwhile, \citet{2011MNRAS.413..183M} proposed that wind Roche-lobe overflow \citep[wind RLOF; based on the mechanism of][]{2007ASPC..372..397M} can provide stable and substantial mass transfer in such systems.
Then, using the arguments of \citet{1989ApJ...346..847C} and \citet{1994ApJ...436..843B} mass can be hypercritically accreted onto the BH to provide the observed spins.

In section~\ref{sec-CCSN} we describe the collapse of the core of massive stars, the formation of a SASI from the original shockwave produced by the bounce of the homologous core and how this may spin up NSs or BHs.
In section~\ref{sec-Model} we describe the model employed to estimate the maximum $a_\star$ possible through a SASI and detail the conservation-laws requirements on the spin-up of central compact objects.
In section~\ref{sec-Res} we present our results, and discuss their implications in section~\ref{sec-Discu}.
We show our conclusions in  section~\ref{sec-Concl}.

\section{Core Collapse and SASI}
\label{sec-CCSN}

When the iron core of a massive star becomes too massive, it can no longer maintain hydrostatic equilibrium.
As the stellar core collapses into a PNS an accretion shock forms.  Based on numerical simulations, \citet{2009ApJ...694..664M} argue that if the shock reaches a radius larger than some $500$ km then it will, most likely, launch a supernova.  
However, before that occurs, the accretion shock will be drained of energy (by photodissociating the infalling Fe into neutrons and protons) and it may stall.

If an accretion shock lingers it may develop instabilities \citep[e.g.,][]{2002A&A...392..353F,2003ApJ...584..971B} and
\citet{2007Natur.445...58B} suggest that a SASI may be formed. 
This instability has a dominant $m=1$ spiral mode (where $m$ is the azimuthal wave number). 
In their numerical modeling it seems the SASI develops preferentially around the original spin vector of the collapsing star \citep[but see also][]{2011ApJ...732...57R}.
As the SASI develops, the wake of the instability wraps over its tail.
The rear part of the shock and a stream of material is sent towards the PNS imparting a torque in the direction opposite to the rotation of the SASI (and the star).
Thus, for as long as the spiral-mode SASI spins around the PNS, the latter will receive a torque and will be spun up into negative values of $a_\star$.
In the context of a binary that survives the SN, negative values of $a_\star$ correspond to a spin vector with direction opposite to that of the accreting material (and due to tidal synchronization, opposite to the orbital angular momentum vector, as we discussed in section~\ref{subsec-KBHs}).

As long as the mass of the hot PNS does not exceed a value of about $2.5 - 3.0~\msun$, it may remain stable against further collapse into a BH.
Strong differential rotation may allow for an extra solar mass
on the (static) maximum mass \citep[][]{2000ApJ...528L..29B,2009CQGra..26f3001O}.
However when accretion drives the PNS mass above some critical mass, a BH is formed.
The exact maximum value for the mass of a PNS (or, for that instance, a NS) depends on the nuclear equation of state (EoS) and is not well established. 
We do know, however, it is above $M^{\rm max}_{\rm NS} > 2 \msun$ from the observations of \citet{2010Natur.467.1081D} and \citet{2013Sci...340..448A}.
At the same time BH masses as small as $3 \msun <\Mbh \lesssim 4 \msun$ are expected (e.g., for GRO J0422+32 \citet{2003ApJ...599.1254G} estimate $\Mbh = 3.97 \pm 0.95 \msun$; also, \citet{1999PASP..111..969F} estimate $3 \msun < \Mbh < 4.8 \msun$ for the BH in XN Vel 93; \citet{2015arXiv150600181G} suggest that V1408 Aquilae may harbour a $3 \msun$ BH). 
Therefore here we will assume the maximum mass for a PNS to be $\Mpns = 3~\msun$.
This value is larger by $\sim0.5~\msun$ than those in \citet{2010arXiv1012.3208L}, but we will show in section~\ref{sec-Res} that our results depend only very weakly on the choice of $\Mpns$.

When the inner PNS or NS collapses into a BH, the pressure support from the  compact object vanishes and a substantial source of neutrinos disappears as well.
The fragile balance allowing the SASI to exist suddenly extinguishes and within a millisecond it is dragged into the event horizon (Foglizzo, 2012, private communication; see also, e.g., the simulations by \citet{2011PhRvL.106p1103O} and Fig.~3 in \citet{2011ApJ...730...70O}).
Thus, the mechanism for spinning the BH up shuts down with the loss of the SASI \citep[see, e.g.,][]{2009CQGra..26f3001O}.

In Sec.~\ref{sec-Model} we discuss a simplified two-stage model for the formation and evolution of a stellar BH.
First, the PNS accretion phase (i.e., the first $3~\msun$), during which the SASI is able to spin up the central compact object.
Second, the BH accretion phase, during which material is accreted from the star but no SASI mechanism is present.
For a SASI to spin up a PNS it is important to keep in mind the following:
\bi
\item{There has to be a source of energy to provide the rotational energy deposited in the core and the envelope of the collapsing core assuming that the progenitor had little angular momentum.}
\item{A SASI must be present.  Once the SASI is disrupted, the mechanism to spin up the PNS shuts down.}
\item{The angular momentum deposited in the core has to be  balanced by angular momentum carried away by material outside the SASIs surface.}
\item{The ratio of rotational kinetic energy ($T$) to gravitational binding energy ($W$) is limited by the onset of instabilities that may trigger the loss of energy through gravitational waves.} 
\ei

\section{The Model}
\label{sec-Model}

We consider the collapse of a massive star  and divide it into two phases. The PNS and the BH phases.
The first stage involves the homologous core collapse and the formation and growth of the hot PNS. This phase ends with the formation of a BH
when the mass of the PNS reaches the critical value of $3~\msun$.
During the PNS phase part of the gravitational binding energy may be used, after homologous-core bounce, to spin the PNS up.
As discussed in Sec.~\ref{sec-CCSN} this may be achieved by forming an instability in the standing accretion shock.
The SASI is driven by the copious neutrino flux coming from the hot material that makes the PNS as well as from the material accreting onto the PNS. 
Thus, we assume that the SASI can only exist while these two neutrino sources are present.

During the BH phase we follow the accretion of the remaining stellar material onto the newly formed BH.
In the case of the HMXBs in Tab.~\ref{Tab:Binaries}, the pre-collapse stars are expected to be spinning  rather slowly \citep{2007ApJ...671L..41B,2008ApJ...689L...9M,2010arXiv1011.4528A,2011MNRAS.413..183M}.
This is because, as mentioned in section~\ref{subsec-KBHs}, such stars should be tidally synchronized and given the massive companions, the orbital separation cannot be too small. 
Moreover \citet{2007Natur.445...58B} have shown that the most likely direction of the spin imparted by the SASI to the central compact object is, preferentially, opposite to any pre-existing angular momentum vector.
For this reason, and with the aim of calculating an upper limit to the possible natal Kerr parameter of BHs in HMXBs, we accrete material with zero angular momentum.
Thus, were $J$ different from zero, it would have opposite direction to $J_{\rm BH}$. 

As already stressed, our model only aims  at establishing an upper limit to the spin rate imparted by the SASI to the underlying compact object.  
With this in mind we assume that the angular momentum provided by the SASI to the PNS is balanced by a ring of material of some mass $\Delta M$ leaving the star with $J_{\rm ring}\!=\!-J_{\rm PNS}$.
A ring is chosen over a sphere as the material close to the poles carries little to no angular momentum, thus it is more efficient in removing angular momentum.
As the PNS collapses into a BH angular momentum is conserved and $J_{\rm PNS} = J_{\rm BH}$.
Since $a_\star\propto J_{\rm BH} \Mbh^{-2}$, fixing $J_{\rm BH}$ to a constant value and increasing $\Mbh$ will rapidly reduce $a_\star$.
Obviously, if any of the mass containing the balancing angular momentum $J_{\rm ring}$ were not to leave the star but eventually accrete into the BH, then $a_\star$ would decrease more rapidly.
If the whole star were to collapse, no angular momentum would escape and the BH should recover the original angular momentum ($J_{\rm star}\sim 0$) of the pre-collapse star.

We further assume that all the binding energy of the collapsed core is available to be converted into rotational energy.
Notice that under these circumstances, since no energy is lost but rather only converted into rotation, the baryonic and the gravitational mass are the same.
We estimate the specific binding energy $E_{\rm B}$ of the collapsing core as:

\begin{equation}\label{eq-E_B}
E_{\rm B}\!=\!1-E\!=\!1\!-\!\frac{1-\frac{2m}{r}\pm a_\star\left(\frac{m}{r}\right)^{3/2}}{\left[1-\frac{3m}{r}\pm2a_\star\left(\frac{m}{r}\right)^{3/2}\right]^{1/2}},
\end{equation}
where $E$ is the specific kinetic energy of a particle orbiting at a distance $r$ from a central compact object with mass  $m\!=\!GM_{\rm PNS}/c^2$ \citep[expressed in length units, see, e.g.,][]{2009blho.book.....R}. 
The maximum binding energy is then obtained imposing the smallest possible radius of an equatorial stable circular orbit. 
This corresponds to the marginally stable orbit $R_{\rm ms}$. 
During hypercritical accretion, which is clearly the case during core collapse 
($\dot{M}_{\rm CC} \simeq 0.1 \msun$ s$^{-1} >> \dot{M}_{\rm Edd}\simeq 10^{-15}\msun$ s$^{-1}$), 
the radius of the smallest orbit of the in-falling material can move inward from the marginally stable to the marginally bound \citep[$R_{\rm mb}$; see e.g.,][]{1978A&A....63..209K}. 
However, the binding energy associated to $R_{\rm mb}$ is zero \citep[see, e.g. Fig.~25.2 in][for the $a_\star\!=\!0$ case]{1973grav.book.....M}. This means that by using $r=R_{\rm ms}$ in Eq.~\ref{eq-E_B} we are extracting the (local) 
maximum possible binding energy from the collapse process, again in the spirit of providing an  upper limit for $a_\star$.

\be
E_{\rm Rot,PNS}\!=\!f(a_\star)\Mpns c^2,
\ee

Energy conservation requires the energy released during core collapse to be greater or equal to the rotational energy acquired by the PNS (which we here denote by $E_{\rm Rot,PNS}$) plus the rotational energy of the ring of material balancing the angular momentum (denoted by $E_{\rm Rot,ring}$), thus
\be
E_{\rm B}\ge E_{\rm Rot,PNS} + E_{\rm Rot,ring}, \label{eq-Econserv}
\ee
where

\be
f(a_\star)=1-\sqrt{\frac{1}{2}\left(1+\sqrt{1-a_\star^2}\right)}. 
\ee

The energy associated with the rotating ring  $E_{\rm Rot,ring}$ should in principle be calculated in general relativity. However, the difference between the special and the general relativistic treatment becomes significant only close to the central compact object. 
The radius of the standing accretion shock where the SASI develops is on the order $R_{\rm SASI} \simeq 150$ km $\simeq 100 R_{\rm Sch}$ \citep[][]{2003ApJ...584..971B,2011ApJ...730...70O}. 
On the other hand the marginally stable orbit for a 3$\msun$ BH (or PNS) is located at 27 km (for $a_\star\!=\!0$) and 4.5 km (for $a_\star\!=\!1$).
As $R_{\rm SASI} >>  R_{\rm ms}$  we can simplify our approach and calculate the energy in the framework of special relativity as:

\be
E_{\rm Rot,ring}\!=\!\frac{1}{2}\,I_{\rm ring}\,\omega^2,
\ee

with moment of inertia

\be
I_{\rm ring}\!=\!\gamma \, \Delta M R_{\rm SASI}^2,
\ee

where $\Delta M$ is  the mass of the ring, 
$\omega\!=\!J_{\rm BH}/I_{\rm ring}$ is the ring angular velocity obtained from angular momentum conservation ($J_{\rm ring}\!=\!-J_{\rm BH}$) and the Lorentz factor
\be
\gamma\!=\!\left( \sqrt{1-\frac{\omega R_{\rm SASI}}{c}} \;\right)^{-1}.
\ee

In Sec.~\ref{sec-Res} we find that $E_{\rm Rot,ring}$ is smaller than $E_{\rm Rot,BH}$. 
However, as the radius and mass of the ring decrease the energy necessary to spin it up (so that angular momentum is conserved) grows rapidly.
This means that the PNS spin is further limited, as more and more of the binding energy goes into spinning up the ring instead of the PNS, as can be seen in Fig.~\ref{fig-logTWvsA}.

It is also important to note here that we assume all the material inside the SASI to be accreted through the equatorial plane.  In fact, eq.~\ref{eq-E_B} is valid for circular equatorial orbits. We argue that this is not a bad assumption as the material interacting with the SASI (either crossing it or forming a part of the ring that carries the balancing angular momentum away) will have large specific angular momentum parallel to the rotational axis and so it will have a tendency to form an equatorial accretion/decretion disk.

\section{Results}
\label{sec-Res}

\subsection{Limits On The Conversion Of Binding Energy Into Rotational Energy}
\label{subsec-Bar}

In our model during the PNS spin up, gravitational binding energy is converted into rotational energy by the SASI.
The ratio of specific kinetic rotational energy of the central compact object to specific gravitational binding energy, $T/|W|$ (where $T$ is the kinetic energy of the PNS and $|W|$ is the available binding energy from stellar collapse), as a function of the Kerr parameter is shown in Fig.~\ref{fig-logTWvsA}.
If we neglect the energy spent on spinning up the counter-rotating ring, the curve in the plot is independent of the mass of the central compact object.  
Then we see that to form a BH with a Kerr parameter as high as 0.2, almost 10\% of the gravitational binding energy has to be converted into rotational energy. 
Since the binding energy is calculated at the marginally stable orbit of a rotating compact object in 
Eq.~\ref{eq-E_B}, we are already considering an upper limit for the available energy. 
In fact, as in our scenario the star is born with little or no angular momentum (see previous sections), the central compact object is formed with $a_\star \sim 0$, and has to spin up gradually. 
If we assume $100\%$ efficiency of energy conversion (from binding to rotational) and we neglect the energy required to spin the counter-rotating ring, the maximum physically achievable spin is $a_\star\!\simeq\!0.94281$ (Fig.~\ref{fig-logTWvsA}).

If another source of energy were available (e.g., nuclear burning), it could, in principle, be possible to cross beyond $a_\star\!\simeq\!0.94281$ and reach a region where the binding energy is, again, substantially larger than the rotational energy.
Nonetheless, we must remember that as material crosses the SASI it drains its energy as Fe gets dissociated into neutrons and protons (as much as $\sim 18$ bethes per $\msun$), and we have not accounted for this.

However, such large efficiencies are, likely, out of reach.
The energy conversion efficiency, as measured by $T/|W|$, 
will be limited by, among other factors, the deformation of the rapidly rotating PNS and the onset of secular and dynamical instabilities that trigger the production of 
gravitational waves \citep[see e.g., the discussion on this topic in][]{1983bhwd.book.....S}.
Secular instabilities are triggered at $T/|W|\!\lesssim\!0.14$ and dynamical instabilities at $T/|W|\!=\!0.27$ \citep{1983bhwd.book.....S}.
Exceeding by a large factor the dynamical-instability critical value of $T/|W|$ excites the bar-mode instability (the PNS deforms into a long 'bar-shaped' ellipsoid, i.e. $m\!=\!2$) 
which leads to gravitational torques produced by spiral arms which rapidly remove the excess angular momentum. 
Thus $T/|W|$ is effectively prevented from growing above such critical values.
This in turn will limit the maximum spin that the PNS, and hence the BH, can obtain from the SASI.
For $T/|W|\!=\!0.14$ the maximum absolute value of the Kerr parameter at the end of the PNS phase would be $|a_\star^{\rm max}|\!=\!0.27$, or, if we were to ignore the secular instabilities, 
at an efficiency $T/|W|\!=\!0.27$ we would obtain $|a_\star^{\rm max}|\!=\!0.38$.
Thus, under our assumptions, $|a_\star|\!=\!0.38$  is a rather solid upper limit, and $|a_\star|\!=\!0.27$ is a conservative upper limit for $|a_\star^{\rm max}|$ at the end of the PNS phase.

One last word of caution with regard to the onset of rotational instabilities.  \citet{2002MNRAS.334L..27S} find that if the PNS has a
large rotational gradient (and it has a stiff EoS, i.e. $P\!=\!K\rho^\Gamma$ with a polytropic index $\Gamma\!=\!2$, where $P$ is the pressure, $\rho$ the density and $K$ is the polytropic constant), dynamical instabilities may occur for $0.03\!\lesssim\!T/|W|\!\lesssim\!0.15$.
Furthermore, \citet{2003MNRAS.343..619S} later find that they can generalize these results for different polytropic indices and differential rotation profiles and obtain dynamical instabilities with $T/|W|$ as low as $0.01$.
Later, \citet{2007PhRvL..98z1101O} \citep[see also][]{2012PhRvD..86b4026O} have performed simulations with realistic EoSs that include microphysics (instead of polytropes). 
They confirm the onset of dynamical instabilities (with $m\!=\!1, 2$ and 3) for $0.01\!\lesssim\!T/|W|\!\lesssim\!0.15$ in the PNS.
Using Fig.~\ref{fig-logTWvsA}, this implies a range $0.07\!\lesssim\!|a_\star^{\rm max}|\lesssim\!0.27$  for the natal Kerr parameter at the beginning of the BH phase.

\subsection{HMXBs: The case of \cyg}
\label{sec-AngMom}

Recently \citet{2011arXiv1106.3688R}, \citet{2011arXiv1106.3689O} and \citet{2011arXiv1106.3690G} have estimated the distance, masses, and spin of \cyg.
They find $a_\star>0.97$ (however see also \citet{2010arXiv1011.4528A} who estimate the spin of the BH in \cyg using quasi-periodic oscillations and find $a_\star=0.48\pm0.01$, or \citet{2009ApJ...697..900M} who find $a_\star\!=\!0.05$).
A spin larger than $a_\star\!\gtrsim\!0.15$ is at odds with theoretical expectations of natal spin for this object  \citep{2011ApJ...727...29M,2007ApJ...671L..41B,2010arXiv1011.4528A}, since
the stellar companion is too large to fit in an orbit that would produce a large $a_\star$ from tidal synchronization (see Tab.~\ref{Tab:Binaries}).
This is similar to the case of \mx  and \lmc, where the expected natal spin of the of the BHs is much smaller than the measured one \citep{2008ApJ...689L...9M,2011MNRAS.413..183M}.
As mentioned in section~\ref{sec-Intro},
\citet{2010arXiv1011.4528A} suggest that the spin of the BH in \cyg may have been acquired through the mechanism put forward by \citet{2007Natur.445...58B}.
To check this possibility we apply here the model discussed in section~\ref{sec-Model} to the central BH of \cyg.

\citet{2003Sci...300.1119M} studied the velocity of \cyg with respect to the Cyg OB3 star association and obtained $V=9\pm2$ km s$^{-1}$.
Thus they deduce that the mass loss during the formation of the BH was little if not zero.
Accordingly, \citet{2010arXiv1011.4528A} choose a BH mass of $\Mbh \!=\!9\msun$ and a mass loss $\Delta M\!<\!1~\msun$.
Following this approach we assume  $\Delta M\!=\!1\msun$ in our calculation.
This is in agreement with the updated measurements by \citet{2011arXiv1106.3688R}, who find a velocity with respect to the Galaxy of $\sim\!21$ km s$^{-1}$, which, when translated to the reference frame of the Cyg OB3 cluster is about $\sim\!5\pm3$ km s$^{-1}$ (Reid, private communication). 

\begin{figure}[h!]
\begin{center}
\includegraphics[width=0.98\columnwidth]{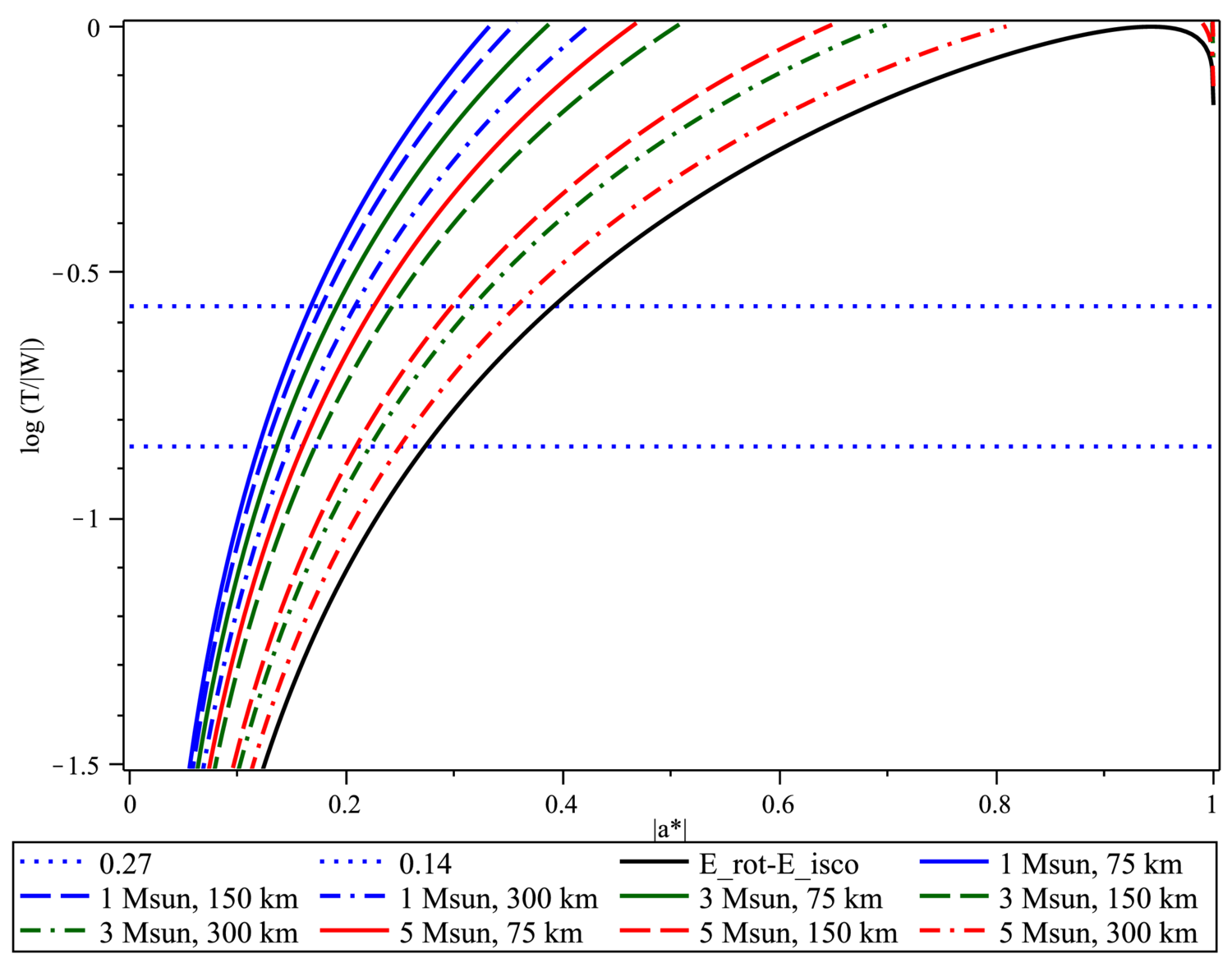}
\caption{Ratio of rotational kinetic energy to gravitational binding energy $\log(T/|W|)$ as function of the Kerr parameter $a_\star$ of the collapsed stellar core at the moment of BH formation.
Also plotted are the limits for the onset of secular ($T/|W|\!=\!0.14$) and dynamical ($T/|W|\!=\!0.27$) instabilities (blue-dotted lines).  
These curves are relevant upto $\Mpns \simeq 3\msun$ as the SASI likely dissapears a few milliseconds after BH formation.
The black curve represents the simple case in which the energy required to spin up the counterrotating disk is neglected (total angular momentum is not conserved).
The blue curves are for a $1\msun$ disk,  green  for a $3\msun$  and  red  for a $5\msun$ disk. 
For each disk mass we show results for disk formation at different radii:   75km (continuous line), 150km (long-dashed line),
 300km (dashed line) and  600km  (dot-dashed line).
For the $1\msun$ disk we  also show the 1200 km radius case (space-dashed line).
\label{fig-logTWvsA} 
}
\end{center}
\end{figure}

The available energy from collapse during the PNS phase is about 2000 bethes (1 bethe = $10^{51}$ ergs).
This energy is enough to spin up the newly formed 3$\msun$ black hole to high values of the Kerr parameter (see Fig.~\ref{fig-logTWvsA}).
Indeed the rotational energy of a maximally rotating (i.e., $a_\star\!=\!1$) 3$\msun$ black hole is about 1600 bethes, with only a few bethes of rotational energy required by the counter-rotating 1$\msun$ ring to conserve angular momentum.
Therefore at the end of the PNS phase the SASI may efficiently spin up a newly formed black hole to high values of the Kerr parameter.
Of course this relies on the assumption of efficient conversion of binding energy into rotational energy.

We now follow the evolution of $a_\star$ after the PNS exceeds $3~\msun$,  collapses into a BH (BH phase) and the SASI dissapears.
We find that any significant amount of mass accreted by the BH rapidly decreases its Kerr parameter.
This is because, as discussed in Sec.~\ref{sec-Model}, while accreting material with little or no angular momentum the spin parameters decreases as $a_\star \propto M_{BH}^{-2}$.
By the time the BH reaches a mass of $\sim10\msun$ we expect $|a_\star|\!<\!0.1$, regardless of how large the Kerr parameter was at the end of the PNS phase (see Fig.~\ref{fig-AvsMbh}).
We conclude that, in the case of \cyg and more in general for the HMXBs listed in Tab.~\ref{Tab:Binaries},
a spiral SASI  can not explain Kerr parameters higher than $|a_\star|\!\sim\!0.1$.
The same should hold true for other IMXBs and HMXBs
with  $\Mbh\!\gtrsim\!6~\msun$ .

One may further ask about what would happen if we remove the assumption that the spin of the SASI aligns with that of the progenitor, producing a counterrotating BH.  
Would the angular momentum from tidal synchronization added to that produced by a SASI be able to explain a considerably larger $a_\star$?
Here one has to assume, again, that the ring would leave the star without removing too much angular momentum from the outer layers.  
To maximize this effect let us assume the final angular-momentum vector of the BH is aligned with that of the binary. 
From its current orbital period, $5.6$ days and considering a maximum mass loss of $1 \msun$ we know the orbital period of the binary at the time of BH formation would be $\sim\!5.3$ days.
From Fig.~3 in \citet{2008ApJ...685.1063B} we see that the $a_\star$ increase would be less than $0.1$ to our estimate above, i.e. $a_\star\!\le\!0.2$.
Nowhere near the $a_\star\!\simeq\!0.5$ of \citet{2010arXiv1011.4528A} and even further from the $a_\star\!\lesssim\!0.97$ of \citet{2011arXiv1106.3690G}.

\subsection{Spin Up Of Low Mass BHs}
\label{subsec-P2}

Let us now consider the case in which a successful SN explosion removes all the material above a massive PNS which eventually, as it cools and contracts, collapses to a BH.
In our simplified picture, this material would carry away angular momentum to balance the formation of a BH with a substantial Kerr parameter (again, from Fig.~\ref{fig-logTWvsA}, $0.27\!\lesssim\!|a_\star^{\rm max}|\!\lesssim\!0.38$ for $0.14\!\lesssim\!T/|W|\!\lesssim\!0.27$).
However such a BH would be a very low mass one.  Later fallback could increase its mass while lowering its $a_\star$.

It is important to note that this result is consistent with \citet{2007Natur.445...58B}, who claim that for NSs substantial Kerr parameters can be acquired through the SASI mechanism.
The $|a_\star|$ of a $1.5~\msun$ PSR with a spin period of $P=10^{-2}$ seconds (similar to the largest measured spin period of a non recycled PSR) is $\sim0.05$. 
This value is well within the limits of what we can obtain even when secular instabilities develop.
In fact, even if we adopt $1\%$ efficiencies, these numbers are much more in line with tipical kinetic 
energies to neutrino luminosity energies, or with those obtained from PNS instabilities such as those estimated in \citet{2003MNRAS.343..619S} and \citet{2007PhRvL..98z1101O}.

\begin{figure}[h!]
\begin{center}
\includegraphics[width=0.98\columnwidth]{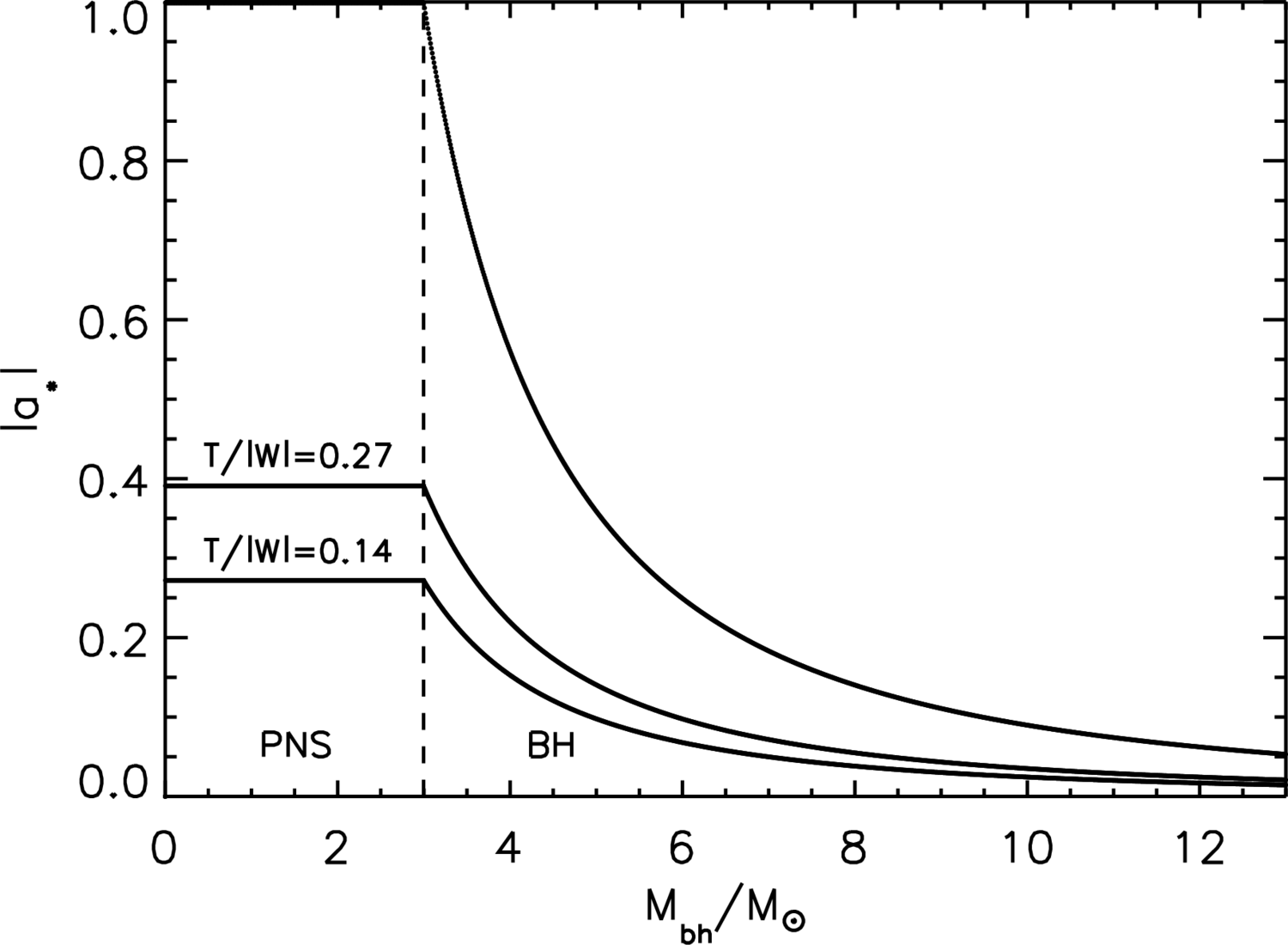}
\caption{Evolution of $a_\star$ as function of mass of the central compact object (in solar masses).  During the PNS/SASI phase gravitational potential energy can be converted into rotational kinetic energy.  However, after BH formation the SASI is not present anymore and the accretion of mass with little or no angular momentum brings down the absolute value of $a_\star$.  The evolution is shown for energy conversion limited by secular instability, $T/|W|\!=\!0.14$, dynamical instability, $T/|W|\!=\!0.27$, and by conservation of energy, $T/|W|\!=\!1$.  In the latter case, we have limited the Kerr parameter to $|a_\star|\!\le\!1$ in agreement with the cosmic censorship conjecture \citep{1969NCimR...1..252P} and with the limits imposed by \citet{1974ApJ...191..507T}. \label{fig-AvsMbh} %
}
\end{center}
\end{figure}

The system  GRO J1655$-$40 contains the least massive BH with an $a_\star$ measurement.
With a Kerr parameter value of $a_\star\!=\!0.65-0.75$, this leaves out the possibility that its spin was obtained through a spiral SASI, as we can see from Fig.~\ref{fig-AvsMbh} that at $\Mbh\!=\!5.4 \msun$ one would not expect a value of $|a_\star|$ anywhere above 0.3 if the material accreted had zero angular momentum and even less if the material had been synchronized with the orbit.
Nevertheless, the spin of the BH in this binary can be very well explained by assuming pre-collapse tidal synchronization and an increase in the orbital period due to mass loss from the SN (the value of $a_\star$ was predicted in \citet{2002ApJ...575..996L} and \citet{2007ApJ...671L..41B}).

\section{Discussion}
\label{sec-Discu}

In the models of \citet{2007Natur.445...58B} the  $m=1$ mode SASI preferentially acquires a similar axis of rotation to that of the collapsing star \citep[however see][]{2011ApJ...732...57R}.
The PNS ends up counter-rotating with regard to the SASI and, roughly, to the initial spin of the star.
The Kerr parameter
in all measured stellar BH binaries (with the possible exception, given the large error bars, of A0620$-$003) seems to be positive (see tables in \citealt{2011arXiv1101.0811M} and \citealt{2011arXiv1102.1500M}), that is the spin and orbital angular momentum vectors point in the same direction. 
This is expected, due to the likelihood of tidal synchronization of BH progenitors prior to collapse and argues against a SASI origin of the spin of  BHs in X-ray binaries, and in particular in HMXBs.
One must keep in mind that tidal synchronization in binaries such as those discussed here will likely produce collapsing stars with $J\!\neq\!0$.
Thus, the spiral SASI will first slow down the PNS to $a_\star\!=\!0$ and later it will spin it up towards negative values of the Kerr parameter.
The absolute values of $a_\star$ at the end of the PNS stage will be likely lower than what we have estimated.
Also, during the BH phase, as more material from the envelope is accreted, $a_\star$ will cross again through its zero value and recover a positive value.
In principle, if no angular momentum were to be removed from the star, one would recover values similar to those obtained in, e.g., \citet{2011MNRAS.413..183M} and \citet{2011ApJ...727...29M}, i.e., $0\!<\!a_\star\!<\!0.15$, corotating with the binary, for \cyg, \lmc and \mx.

An important energy-draining channel that could limit the efficiency of the SASI mechanism is the appearance of strong magnetic fields.
\citet{2010AAS...21642804A} suggests that the non-axisymmetric mode of the spherical accretion shock instability  creates strong shear.
This in turn can significantly amplify a pre-existing magnetic field.
Such magnetic field amplification could lead to stronger angular momentum transport, limiting further the spin-up efficiency  of the SASI.
The possibility that this could lead to the formation of  a fast rotating, strongly magnetized, compact object is intriguing \citep{Cheng:2014}. Such exotic remnants, under specific conditions, are believed to be the central engines of Long Gamma-Ray Bursts \citep[e.g,][]{1994MNRAS.270..480T,2000PhR...325...83L,2008MNRAS.385L..28B}.

\section{Conclusions}
\label{sec-Concl}

We studied the role of a non-axisymmetric mode of the spherical accretion shock instability (SASI) in spinning up core collapse remnants.
Assuming the SASI is the main driver of the spin-up process, our results predict a strong dichotomy in the maximum spin of low mass compact objects and massive BHs found in HMXBs.
This mechanism can in principle result in substantially large natal spin of low mass compact objects (either NSs or BHs). The maximum value of $a_\star$ for a compact object near the boundary between BHs and NSs is found to be somewhere between 0.27 and 0.38, depending on whether secular or dynamical instabilities limit the efficiency of the spin up process.
On the other hand, our results show that for the more massive BHs found in HMXBs, the natal spin is substantially smaller.
This is because any initial spin-up imparted by the SASI mechanism is reduced by further accretion
of low specific angular momentum material
during the stellar collapse. 
In particular we found that for $\Mbh\!>\!8~\msun$  the  SASI  mechanism  can only produce values of the Kerr parameter $a_\star\!\lesssim\!0.05$. Even neglecting the limits imposed by dynamical instabilities in a rotating compact object, we obtain $a_\star\!\lesssim\!0.15$.
Therefore we conclude that  the observed Kerr parameters of BHs in HMXBs such as \mx, \lmc or \cyg can not be the result of a spiral SASI spin up.
In these cases hypercritical accretion may be necessary to spin up these BHs to the observed values, as proposed in \citet{2011MNRAS.413..183M}.

\section*{Acknowledgements}
\label{sec-Ack}

We thank M. Liebend\"{o}rfer, C. Ott, T. Foglizzo and F. Lora-Clavijo for useful discussions.  
EMM was supported by a CONACyT fellowship and projects CB-2007/83254 and CB-2008/101958.  
This research has made use of NASA{'}s Astrophysics Data System as well as arXiv.



\bibliographystyle{elsarticle-num}



\end{document}